\documentclass[aps,prl,twocolumn,noshowpacs,superscriptaddress]{revtex4-2}

\usepackage{graphicx}% Include ure files
\usepackage{dcolumn}% Align table columns on decimal point
\usepackage{bm}% bold math
\usepackage{mathtools}
\usepackage{xcolor}
\usepackage{layouts}
\usepackage{hyperref}
\usepackage{amssymb}
\usepackage{dsfont}
\usepackage{caption}
\usepackage{subcaption}

\usepackage{blindtext}

\renewcommand{\P}{\mathcal{P}}
\renewcommand{\S}{\mathcal{S}}

\newcommand{\affilLAN}{Department of Physics, Lancaster University, Lancaster LA1 4YB, United Kingdom.}

\begin{document}

\preprint{APS/123-QED}

\title{Action formalism for geometric phases from self-closing quantum trajectories}

\author{Dominic Shea}\affiliation{\affilLAN}
\author{Alessandro Romito}\affiliation{\affilLAN}

\date{\today}

\begin{abstract}
When subject to measurements, quantum systems evolve along stochastic quantum trajectories that can be naturally equipped with a geometric phase observable via a post-selection in a final projective measurement. When post-selecting the trajectories to form a close loop, the geometric phase undergoes a topological transition driven by the measurement strength. Here, we study the geometric phase of a subset of self-closing trajectories induced by a continuous Gaussian measurement of a single qubit system. We utilize a stochastic path integral that enables the analysis of rare self-closing events using action methods and develop the formalism to incorporate the measurement-induced geometric phase therein. We show that the geometric phase of the most likely trajectories undergoes a topological transition for self-closing trajectories as a function of the measurement strength parameter. Moreover, the inclusion of Gaussian corrections in the vicinity of the most probable self-closing trajectory quantitatively changes the transition point in agreement with results from numerical simulations of the full set of quantum trajectories.
\end{abstract}

\maketitle

\section{Introduction}

The global phase of a system’s quantum state is an unmeasurable $U(1)$ gauge freedom. 
However, when a system is driven adiabatically in a closed cycle, the accumulated phase difference becomes gauge invariant and, therefore, experimentally accessible~\cite{Atala_2013, doi:10.1126/sciadv.aay2730}. 
This observable phase difference consists of two components: a dynamical (or local) phase and a Berry (or geometric) phase~\cite{berry1984quantal, PhysRevLett.51.2167}, which is solely dependent on the path taken through the Hamiltonian's parameter space.
Geometric phases are key to identifying topological phases of matter like quantum Hall phases, topological insulators and superconductors~\cite{chruscinski2012geometric,RevModPhys.82.1959,Asb_th_2016, RevModPhys.80.1083}, and have been exploited in the design of high-fidelity quantum gates that are resilient to random noise~\cite{PhysRevA.72.020301, Sj_qvist_2015}.

Geometric phases can be generalised to continuous evolution in a system's Hilbert space via the Aharonov-Anandan phase, which can be defined as a holonomy element of a fibre bundle over the projective Hilbert space of the system~\cite{chruscinski2012geometric, PhysRevLett.58.1593,2019NatPh..15..665C}.
Avoiding the use of Hamiltonian parameter space, this approach places no restrictions on the system dynamics, so that a geometric phase can be associated with continuous non-unitary dynamics due to quantum measurement back-action~\cite{jacobs_2014,Jacobs_2006}. 
Importantly for measurement-induced effects, the dynamics acquire stochastic properties. 
The difference between the average state and the full distribution of states along quantum trajectories leads to new features in single-qubit dynamics~\cite{PhysRevResearch.2.033512,PhysRevResearch.2.043420} and new out-of-equilibrium many-body states~\cite{Romito2023,PhysRevB.98.205136,PhysRevB.99.224307,PhysRevX.9.031009,PhysRevB.100.064204,Koh_2023,2023,Fisher_2023}. 
Much research has focused on associating some generalized version of the geometric phase to the averaged (mixed) state described by a density matrix, via the Uhlmann phase~\cite{UHLMANN1986229,uhlmann1989berry, uhlmann1991gauge}, or introducing an interferometric geometric phase~\cite{PhysRevLett.85.2845, PhysRevLett.90.160402,PhysRevA.67.020101}.
It is only recently that the statistical properties of the geometric phase along individual stochastic realizations (aka quantum trajectories) have begun to be investigated~\cite{gebhart2019measurement,viotti2023geometric, PhysRevResearch.3.043045,viotti2023geometric,Snizhko_2021,Snizhko_2021_2}.
Recently it has been shown that geometric phases arising from a particular measurement post-selection exhibit a topological transition as a function of the measurement strength~\cite{gebhart2019measurement, PhysRevResearch.3.043045,PhysRevX.9.031009}. 
This transition differs from 
 the topological order transition induced by measurements in many-body systems ~\cite{10.21468/SciPostPhys.14.3.031,PhysRevResearch.2.033347,PhysRevB.102.094204,Lavasani_2021,Wang_2023}, and was originally predicted for a qubit subject to a sequence of cyclically rotating measurements. 
Further work has shown that the topological nature of the transition is a more generic feature of single qubit measurement-induced dynamics including different measurement procedures and protocols, as well as Hamiltonian dynamics and dephasing~\cite{viotti2023geometric,Snizhko_2021,Snizhko_2021_2}. The transition is also robust against several perturbations, including additional adiabatic dynamics and dephasing~\cite{PhysRevLett.127.170401,PhysRevResearch.4.023179}.
This, in turn, has facilitated the experimental observation in superconducting and optical architectures~\cite{PhysRevResearch.4.023179,ferrergarcia2022topological}.

Notwithstanding the various generalizations, the observation of measurement-induced geometric phases requires quantum trajectories that form a closed loop. 
For a generic quantum evolution, this may be achieved with a final post-selection of a projective measurement onto the initial state. While this procedure provides a way to associate a geometric phase to a generic quantum trajectory (open geometric phase hereafter), it introduces a discontinuous jump in the quantum evolution. 
Alternatively, one can define a post-selection procedure of the subset of self-closing quantum trajectories originating from the quantum dynamics, which gives rise to continuous quantum trajectories. 

Here we take advantage of this continuous evolution to develop a path integral formulation of the measurement-induced closed geometric phases, associated with the set of self-closing quantum trajectories. 
In particular, we will follow the approach developed by Chantasri-Dressel-Jordan (CDJ) for continuous Gaussian measurements~\cite{Chantasri_2013, Chantasri_2015,Murch_2013}, which is a well-established technique for investigating continuous measurement dynamics~\cite{Lewalle_2017,Lewalle_2018,Chantasri_2018}.
By incorporating a phase variable in the CDJ action for continuous Gaussian measurements we obtain a path-integral formulation for open and closed geometric phases.
We systematically investigate the most likely open and self-closing quantum trajectories and their associated geometric phase.
We find that the open geometric phase transition remains topological for a variety of post-selection conditions of the readout record and initial state preparations. 
A suitable protocol can be designed to investigate the topological properties of closed-geometric phases. 
In this case, a transition as a function of the measurement strength exists and maintains its topological properties, although the underlying mechanism of competing minima of the probability distribution differs from its counterpart in the open-geometric phases. 
Finally, we develop a Gaussian approximation around the optimal trajectory, which is shown to wapture the statistics of geometric phases resulting from the full distribution of quantum trajectories obtained from numerical simulations.

\section{Geometric Phases Induced by Gaussian Measurements}
\begin{figure}[h!]
\begin{subfigure}{0.5 \textwidth}
\caption{}
\includegraphics[width=7cm]{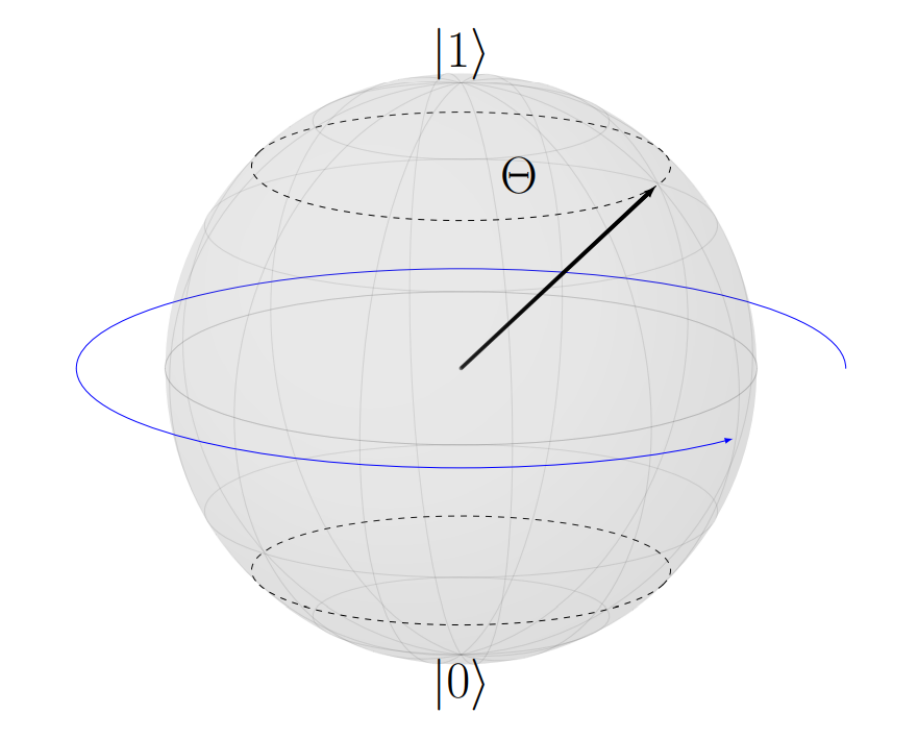}
\label{fig:windingprotocol}
\end{subfigure}
\begin{subfigure}{0.5 \textwidth}
\caption{}
\includegraphics[width=4.75cm]{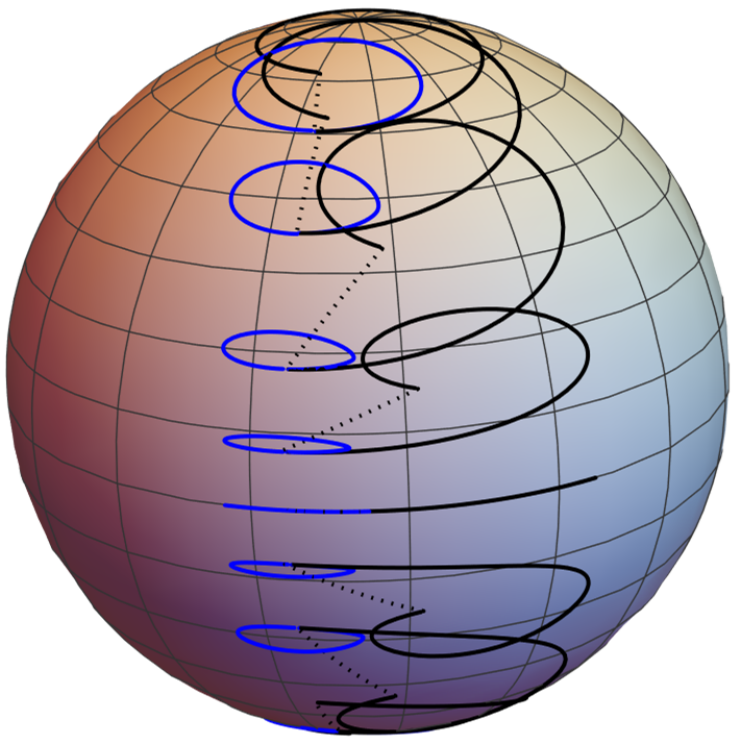}
\label{fig:continuoustrajectories}
\end{subfigure}
\caption{\textbf{Cyclically Rotating Qubit Measurement Protocol and corresponding quantum trajectories}. Panel a) A continuous sequence of Gaussian measurements of operators $\mathbf{\sigma} \cdot \mathbf{n}(t)$ at constant latitude $\Theta$, represented by $\mathbf{n}(t)$ tracing out a path on the Bloch Sphere (dotted lines). Panel b) Examples of quantum Trajectories on the Bloch sphere generated by the measurement sequence in panel (a) for self-closing (blue) and open (black) boundary conditions. Dotted lines indicate the length minimising geodesics closing the open path trajectories.}
\label{fig:measurement axis}
\end{figure}

\subsection{Cyclically rotating Gaussian measurements}\label{MeasurementProtocol}

We consider the measurement-induced time evolution of a qubit, with a generic state given by the density matrix $\rho=\frac{1}{2}\left( \mathbb{I} + \boldsymbol{x} \cdot \boldsymbol{\sigma} \right)$. 
The vector, $\boldsymbol{\sigma}$, consists of Pauli Matrices $(\sigma_X, \sigma_Y, \sigma_Z)$ and $\boldsymbol{x}$ is a unit vector on the Bloch sphere of the system parameterised by latitude $\theta$ and longitude $\phi$.
The qubit is subject to a time-dependent sequence of weak measurements over a fixed time, $T$. 
We define a continuous process by dividing $T$ into $N$ time steps of length $\delta t$, where $t_k = k \delta t $ and $k \in \{n \in \mathbb{N} \cup \{0\} \ | \ n < N \}$. 
During each time interval, we measure the operator $A(k)= \boldsymbol{n}(k) \cdot \boldsymbol{\sigma}$, where $\boldsymbol{n}$ is a unit vector specified by spherical coordinates $(\Theta, \Phi)$. 
We further specify $\Theta(k) = \Theta$ and $\Phi(k) = 2 \pi t_k/T$, so that the target observables are constrained to closed loops of constant latitude on the Bloch sphere (see Fig. \ref{fig:windingprotocol}). 
At each time step, the measurement of the observable $A(k)$ is performed as a Gaussian measurement, with measurement outcomes distributed according to a Gaussian probability distribution~\cite{jacobs_2014, Jacobs_2006}. 

For a system in a state $\rho_k$, the Gaussian measurement of $A_k$ entails a measurement outcome $r_k$ drawn from the probability distribution $P(r_k)= \mathrm{Tr}[E(r_k)^\dagger E(r_k)\rho_k]$ and a corresponding state update $\rho_{k+1}$ given by 
\begin{equation}
P(r_k)= \mathrm{Tr}[E(r_k,k)^\dagger E(r_k,k)\rho_k],
\end{equation}
\begin{equation}
\rho_{k + 1}= E(r_k,k)\rho_k E(r_k,k)^\dagger/P(r_k).
\label{eq:kraus101}
\end{equation}
The entire process is controlled by the set of Kraus operators $E(r_k,k)$. 
For the specific protocol at hand, with measurements constrained at a fixed latitude, we have
\begin{equation}
\label{eq:gauss_simple}
 E(r_k, k)=R(\mathbf{n}(k))^{-1} M_{\delta t}(r_k) R(\mathbf{n}(k)),
 \end{equation}
where
\begin{equation} M_{\delta t}(r_k) = \sqrt[4]{\frac{\delta t}{2 \pi \tau}} \exp\left(-\frac{\delta t}{4 \tau} (r_k-\sigma_{z})^2\right),
\end{equation}
and $R(\boldsymbol{n}(k))$ is a rotation operator $R(\boldsymbol{n}(k)) = e^{-\frac{1}{2}i \phi_k } \text{exp}\left[\frac{1}{2} i (\theta_k  \sigma_2+\phi_k \sigma_3) \right]$ that takes the Block sphere state $(\theta,\phi)$ to $|0 \rangle$.
Here, $\boldsymbol{n}(k)$ denotes the measurement axis at time $t_k$. 

This measurement process is understood by noting that $M_{\delta t}$ corresponds to a Gaussian measurement of the operator $\sigma_Z$. 
The rotational component of $E(r_k, k)$ ensures that the measured operator is $\boldsymbol{\sigma} \cdot \boldsymbol{n}$. 
The parameter $\tau$ sets the characteristic measurement time for each individual Kraus operator, i.e. the time scale after which the state is close to an eigenvalue of the measured observable, or, equivalently, the inverse measurement strength. 
The continuous limit $\delta t \to 0$, $N \to \infty$ with $\delta t N = T$ transforms the sequences of measurement readouts $r_k$ and the corresponding qubit state variables $\phi_k$ and $\theta_k$, which are parameterizations of the qubit state in the spherical Bloch sphere, into continuous functions of time, $r(t)$, $\phi(t)$, $\theta(t)$ defining a continuous stochastic process which we study throughout this paper (see Fig. \ref{fig:continuoustrajectories}).

This measurement protocol is akin to the one presented in Ref.~\cite{gebhart2019measurement}, in which a topological transition in the measurement-induced geometric phase was originally identified. However, a key difference is the use of the Gaussian measurement protocol.  The original protocol involves a quasi-continuous sequence of  measurements with binary outcomes $r(t)=r^j=j$ for $j\in\{1,0\}$ and corresponding Kraus operators $E_j$ defined via Eq. \ref{eq:gauss_simple} with $M_{\delta t} (r_k)$ replaced by $M_j$,
\begin{equation}
    M_{1} = \left( \begin{array}{cc} 1 & 0 \\ 0 & \sqrt{1-\frac{4 c \delta t}{T}}\end{array} \right) \text{and } M_{0} = \left(\begin{array}{cc} 0 & 0 \\ 0 & \sqrt{\frac{4 c \delta t}{T}}\end{array}\right),
    \label{eq:null_kraus}
\end{equation}
where $c$ is a dimensionless measurement strength parameter. 
This choice of Kraus operators replaces the continuous set of Kraus operators parametrized by $r$ in Eq. \ref{eq:gauss_simple}.
In the continuous limit $\delta t \to 0$, The Gaussian measurement backaction  $E(r_k,\mathbf{n}(k))$ continuously approaches the identity operator $\mathbb{I}$, while Eq. \ref{eq:null_kraus} leads to discontinuous jumps associated with the outcome $0$, though with vanishing probability for $\delta t \to 0$.
A quasi-continuous evolution from Eq. \ref{eq:null_kraus} is then distilled by performing a post-selection for `Null' type readout $j=1$. 

\subsection{Geometric Phase of the monitored qubit}

Any of the measurement readout sequences $r(t)$, has an associated trajectory of states on the Bloch sphere $\vert \psi(t) \rangle$. 
Each such trajectory is in turn associated with a unique geometric phase via the functional, \begin{equation}\label{eq:geomphaseFunctional}
    \chi^{\text{g}}[\psi(t)] = \text{arg} \langle \psi(0) | \psi(T) \rangle + i \int_0^T \langle \psi(t) | \dot{\psi}(t) \rangle dt,
\end{equation} where $\vert \psi(t)  \rangle$ is a lift of the curve in projective space into Hilbert space that satisfies an initial condition $\vert \psi(0)  \rangle$ written in some arbitrarily chosen gauge. 
While geometric phases are typically associated with closed paths in Hilbert space, we note that Eq. \ref{eq:geomphaseFunctional} is valid for paths in projective space that do not close~\cite{Sj_qvist_2015}. 
In these cases, the geometric phase is defined as the geometric phase of the given curve in projective space concatenated with a length-minimizing geodesic that closes the trajectory.
We refer to the geometric phase as the \textit{open geometric phase}, as opposed to the \textit{closed geometric phase} associated with self-closing trajectories.
In practical terms, the open geometric phase corresponds to the closed geometric phase of a post-selected trajectory including an additional projective measurement onto $|\psi(0) \rangle \langle \psi(0) |$.

An important feature of Gaussian measurement is that it parallel transports the state of the system, so there is no dynamical phase contribution to subtract from the global phase.
Adopting a pure qubit state parameterisation in spherical coordinates ($\mathbf{q} = (\phi,\theta,\chi)$) that includes a gauge-dependent global phase $\chi$, the systems state is,
\begin{equation} \label{eq:spherical param}
| \psi (\mathbf{q}(t))\rangle  =e^{i \chi(t)}  \Bigg( \begin{array}{c}
\cos \left(\frac{\theta(t)}{2}\right) \\
 e^{i \phi(t)} \sin \left(\frac{\theta(t)}{2}\right) \\
\end{array}
\Bigg).
\end{equation} 
The parallel transport condition $\langle \psi(t+\delta t) | \psi(t) \rangle > 0$, holds across each time step, simplifying the geometric phase from Eq. \ref{eq:geomphaseFunctional} to

\begin{align}
\nonumber \chi^{\text{g}} &= \text{arg}\Bigg[ e^{-i (\phi (0)-\chi (T))} \big(\sin \frac{\theta (0)}{2} e^{i \phi (T)} \sin \frac{\theta (T)}{2} \\ &+ e^{i \phi (0)} \cos \frac{\theta (0)}{2} \cos \frac{\theta (T)}{2} \big)\Bigg],
\end{align}
where we have used gauge freedom to choose $\chi(0) = 0$. 
When $\phi(0)=\phi(T)$ and $\theta(0)=\theta(T)$ then $ \chi^{\text{g}}= \chi(T)$. 

\section{\label{sec:level1}Path Integral: \protect incorporating phase data}

\subsection{\label{sec:citeref} Chantasri-Dressel-Jordan path Integral with Phase tracking}

To study the geometric phases associated with quantum trajectories from Gaussian measurements,
we formulate a path integral for the probability distribution of the induced quantum trajectories that explicitly incorporates the phase of the monitored state. 
We do so by incorporating phase information in the CDJ path integral formulation for a Gaussian-monitored qubit~\cite{Chantasri_2013}, using the state parameterisation in Eq. \ref{eq:spherical param}. 
Since this phase is parallel transported, the global phase is directly equivalent to the geometric phase on a closed path.
The path integral is constructed from the joint conditional probability of finding a state $\mathbf{q}(T)$ and readout $r(T)$, given some initialization $\mathbf{q}_i$, under the evolution in Eq. \ref{eq:kraus101}. 
This probability, in the continuum limit, can be expressed as a product of sequential conditional probabilities so 

\begin{align}
&\mathcal{P}(\mathbf{q}(T), r(T)|\mathbf{q}_i) = \mathcal{F}(t_0,t_N) \times \\ \nonumber  &\lim_{\delta t \to 0, N \to \infty} \prod_{k = 0}^{N - 1} P(\mathbf{q}(k+1) | \mathbf{q}(k), r(k)) P(r(k) | \mathbf{q}(k)),
\end{align}\label{eq:basic_CDJ}
where $\mathcal{F}(t_0,t_N)=\delta^3(\mathbf{q}(N)-\mathbf{q}_f) \delta^3(\mathbf{q}(0)-\mathbf{q}_i)$ sets the initial and final states to be $\mathbf{q}_i$ and $\mathbf{q}_f$. The readout probability distribution function, $P(r | \mathbf{q})$, can be obtained directly from Eq. \ref{eq:kraus101} and \ref{eq:gauss_simple}, so
\begin{equation}\label{eq:conditionalprob}
P(r | q) \approx \sqrt{\frac{\delta t }{2 \pi \tau}} \exp \left( -\frac{\delta t}{2 \tau } (r^2 -2 r a +1) \right),
\end{equation}
\begin{equation}\label{eq:meanreadout}
    a(\Theta,\Phi,\theta, \phi) = \cos (\theta) \cos (\Theta) +\sin (\theta) \sin (\Theta) \cos (\phi-\Phi).
\end{equation}
Here $\langle r \rangle  = \int r P(r|\bold{q}) dr =a$ is the mean measurement record function, and corresponds to the most probable readout at each time step. The remaining term in Eq. \ref{eq:basic_CDJ} is the state update, which is deterministic and can be expressed as a delta function  
\begin{align}
 P(\mathbf{q}(t + \delta t) | \mathbf{q}(t), r(t))&= \delta^d(\mathbf{q}(t + \delta t) | \mathbf{q}(t), r(t)) \\ \nonumber &=\Big(\frac{1}{2\pi i}\Big)^d \int^{i \infty}_{-i \infty} d^d \mathbf{p} e^{-\mathbf{p}\cdot(\mathbf{q}(t + \delta t))},
\end{align}
where by expressing the Dirac-delta functions in Fourier form each $q$ gains a conjugate momentum $p_q$, so $\mathbf{p}=(p_\phi, p_\theta, p_\chi)$. The deterministic update associated with $\delta^d(\mathbf{q}(t + \delta t) | \mathbf{q}(t), r(t))$ is expressed in the differential equations,
\begin{eqnarray}\label{eq:SMEspherical1}
    \dot{\phi} &=  -\frac{r}{\tau } f_{\Theta,\Phi}(\theta, \phi)\\ \label{eq:SMEspherical2}
     \dot{\theta} &= \frac{r}{\tau } g_{\Theta,\Phi}(\theta, \phi,) \\ \label{eq:SMEspherical3}
     \dot{\chi} &= \frac{r}{2 \tau } h_{\Theta,\Phi}(\theta, \phi,).
\end{eqnarray}
with $f_{\Theta,\Phi}(\theta,\phi) =  \csc (\theta) \sin (\Theta) \sin (\phi-\Phi)$,
$g_{\Theta,\Phi}(\theta,\phi)= (\cos (\theta) \sin (\Theta) \cos (\phi-\Phi)-\sin (\theta) \cos (\Theta))$ and $h_{\Theta,\Phi}(\theta,\phi) =\tan \left(\frac{\theta}{2}\right) \sin (\Theta) \sin (\phi-\Phi)$. Following the usual procedure for constructing path integrals~\cite{kleinert2009path,altland_simons_2010,Chantasri_2013,Chantasri_2015,Chantasri_2016}, we express $\mathcal{P}$ in terms of an action principle, 
\begin{equation}\label{eq:path}
    \mathcal{P} \propto \int D\mathbf{q} D\mathbf{p} Dr \exp\left(-\int_0^T S[\mathbf{q},\mathbf{p}, r] dt \right) ,
\end{equation}
\begin{align}\label{eq:StochasticAction}
 S[\mathbf{q},\mathbf{p}, r] =  -p_\theta \left(\dot{\theta} -\frac{r}{\tau } g_{\Theta,\Phi} \right) -p_\phi \left(\frac{r}{\tau } f_{\Theta,\Phi}+\dot{\phi}\right) \\ \nonumber -p_\chi \left(\dot{\chi}-\frac{r}{2 \tau } h_{\Theta,\Phi}\right)  +\frac{r (2 a-r)-1}{2 \tau }.
\end{align}

The action $S[\mathbf{q},\mathbf{p}, r]$, is a generalization of the probability density of a quantum trajectory, encoding all the statistical details of the measurement process. 
Compared to the CDJ action in Ref.~\cite{Chantasri_2013}, Eq. \ref{eq:StochasticAction} now includes the variable $\chi$ and the time dependence of the measurement operators. 
This formalism allows us to identify the most probable trajectories (with a given initial and final state) by variational methods. This results in a system of differential equations for quantum trajectories that are either the most-likely path to traverse between the boundary points or a minimum/saddle point solution, given by Eq. \ref{eq:SMEspherical1},\ref{eq:SMEspherical2} and\ref{eq:SMEspherical3} in combination with the time derivatives of the momentum variables, $\dot{p_\chi} = 0$,
\begin{align}\label{eq:optimumphasespace}
\nonumber \dot{p_\phi} = \frac{r}{\tau} p_\phi \frac{\partial f_{\Theta,\Phi}}{\partial \phi}-\frac{r}{\tau }p_\theta \frac{\partial g_{\Theta,\Phi}}{\partial \phi}-\frac{r}{2\tau }p_\chi \frac{\partial h_{\Theta,\Phi}}{\partial \phi}-\frac{r}{\tau}\frac{\partial a}{\partial \phi}, \\
\dot{p_\theta} = \frac{r}{\tau} p_\phi \frac{\partial f_{\Theta,\Phi}}{\partial \theta}-\frac{r}{\tau }p_\theta \frac{\partial g_{\Theta,\Phi}}{\partial \theta}-\frac{r}{2\tau }p_\chi \frac{\partial h_{\Theta,\Phi}}{\partial \theta}-\frac{r}{\tau}\frac{\partial a}{\partial \theta},
\end{align}
and a constraint on the measurement record function
\begin{align}
r &= \frac{1}{2} p_\chi h(\theta,\Theta,\phi,\Phi) - p_\phi f(\theta,\Theta,\phi,\Phi) \\ \nonumber &+ p_\theta g(\theta, \Theta, \phi,\Phi) + a(\theta,\Theta,\phi,\Phi).
\end{align}
We observe that by introducing the variable $\chi$, the calculation of the optimal geometric phase becomes more straightforward, as its value can be determined concurrently with the optimal quantum trajectory.

\subsection{Corotating coordinates}

It will be useful to consider the action in Eq. \ref{action} rewritten in a spherical coordinate system that co-rotates with the measurement axis defined by new polar and azimuthal coordinates $(\tilde{\theta}, \tilde{\phi})$. 
Such a coordinate transformation is implemented by the change-of-basis matrix $B=\exp(i \Phi \sigma_z/2)$. 
In this coordinate system, the dynamics are described by the Kraus operator $\tilde{E}$ and an effective Hamiltonian $\tilde{H}$. 
Here  $\tilde{E}= R^{-1}(\tilde{\boldsymbol{n}}) M_{\delta t} R(\tilde{\boldsymbol{n}})$, and is now time-independent since $\tilde{\boldsymbol{n}} = (\sin(\Theta),0,\cos(\Theta))$ is fixed. 
While $\tilde{H} = i \dot{B} B^{\dagger} = -\frac{1}{2}  \dot{\Phi} \sigma_z$, and acts unitarily on the system state. 
The state update is given by $\rho(t+\delta t)= \frac{e^{-i \tilde{H}}\tilde{E}\rho\tilde{E}^\dagger e^{i \tilde{H}}}{Tr{e^{-i \tilde{H}}\tilde{E}\rho\tilde{E}^\dagger e^{i \tilde{H}}}}.$
In terms of the action, this point transformation acts directly on the Bloch sphere coordinates as,
\begin{align}\label{eq:co-rot}
\tilde{\phi} &= \Phi - \phi (t) , \quad \tilde{\theta} = \theta,\nonumber\\
    \dot{\phi} &= \frac{\partial \tilde{\phi}(\phi,t)}{\partial \phi}\dot{\tilde{\phi}} +  \frac{\partial \tilde{\phi}(\phi,t)}{\partial t}=-\dot{\tilde{\phi}} + \dot{\Phi}, \quad \dot{\tilde{\theta}} = \dot{\theta},
\end{align}
and also induces a contra-variant transformation on the conjugate momentum,
\begin{equation}
p_{\phi} = \frac{\partial \tilde{\phi}(\phi,t)}{\partial \phi} \tilde{p}_{\phi} = - \tilde{p}_{\phi},  \quad p_{\phi} = \tilde{p_{\phi}}.
\end{equation}
These coordinate transformations applied to Eq.\ref{eq:StochasticAction} (omitting $\chi$ and $p_\chi$), lead to the CDJ action in rotating coordinates,
\begin{align}\label{eq:rotating action}
\nonumber &\tilde{S}[\tilde{\theta},\tilde{\phi},p_{\tilde{\theta}},p_{\tilde{\phi}}] = \frac{1}{2 \tau } \Bigg[+2 r \tilde{a}-r^2-1  \\ \nonumber &+2 \tilde{p}_\theta \Big(r\cos(\tilde{\theta}) \sin(\Theta) \cos (\tilde{\phi})-r \sin(\tilde{\theta}) \cos(\Theta)-\tau  \dot{\tilde{\theta}}\Big) \\  &+2 \tilde{p}_\phi \left(\tau  (\dot{\Phi}-\dot{\tilde{\phi}})-r \csc(\tilde{\theta}) \sin(\Theta) \sin(\tilde{\phi})\right) \Bigg].
\end{align}
with $\tilde{a} =\sin(\tilde{\theta}) \sin(\Theta) \cos(\tilde{\phi})+\cos(\tilde{\theta}) \cos(\Theta)$. From this reformulation, it is evident that the rotating measurement protocol equivalently captures the physics of the Zeno effect, i.e. the competition between measurement and unitary evolution in a qubit.

\subsection{Lagrangian Formulation}

Since the path integral in Eq. \ref{eq:path} is Gaussian in $r$ it is possible to integrate out the measurement record. The ensuing action is quadratic in the momentum variables, which can then also be integrated out to give a configuration space path integral,
\begin{align}\label{eq:pathLag}
\mathcal{P} &\propto \int D\mathbf{q} D\mathbf{p} Dr e(-S[\mathbf{q},\mathbf{p}, r]) \\ \nonumber &= \int D\mathbf{q} 
     \mathfrak{p}[\mathbf{q}] = \int D\mathbf{q} 
     \mu(\theta, \phi) e^{\int_0^T \mathcal{L}[\theta,\phi]}.
\end{align}
This alternate formulation is characterized by a probability measure denoted as  $\mathfrak{p}[\theta, \phi]$ which consists of two components: a singular~\cite{rothe2010classical} Lagrangian $\mathcal{L}[\theta,\phi]$, and a path-dependent functional measure, $\mu(\theta, \phi)$, given by 
\begin{eqnarray} \label{eq:probabilitydensityphi}
    \mu(\theta, \phi)=
    \text{Det}[\frac{\csc ^2(\theta) \sin ^2(\Theta) \sin ^2(\phi-\Phi)}{2 \tau }]^{-\frac{1}{2}},
\end{eqnarray}
\begin{align}\label{eq:lagrangian}
   \nonumber \mathcal{L}(\theta, \phi)  = -\frac{1}{2} \tau  \sin ^2(\theta) \csc ^2(\Theta) \dot{\phi}^2 \csc ^2(\phi-\Phi) -\frac{1}{2 \tau} \\-\frac{\dot{\phi}}{2} \sin (2 \theta ) \cot (\Theta) \csc (\phi-\Phi)-\sin ^2(\theta) \cot (\phi-\Phi) \dot{\phi}.
\end{align}
The path dependence of $\mu$  can be attributed to the multiplicative nature of the underlying stochastic process which acts like a curvature effect in the time axis~\cite{PhysRevE.99.032125}. The Lagrangian is characterised by a non-invertible Hessian matrix. 
This particularity results from the imposition of semi-holonomic constraints within the configuration space, as specified by the equations,
\begin{align}\label{eq:nonholonom}
\dot{\theta}=&\frac{\dot{\phi}}{2} \Big((2 \sin ^2(\theta) \cot (\theta_0) \csc (\phi-\phi_0)\\ \nonumber &-\sin (2 \theta) \cot (\phi -\phi_0)\Big),
\end{align}
\begin{equation}\label{eq:phase constraint}
\dot{\chi} = \frac{\dot{\phi}}{2} (\cos (\theta)-1) .
\end{equation}
These constraints naturally appear during the process of functional integration,  wherein terms exhibiting, at most, linear dependence on momentum play the role of Legendre multipliers. These multipliers, in turn, enforce Eq. \ref{eq:nonholonom} and \ref{eq:phase constraint}.

\section{Topological features of the open geometric phase}

The action (Eq. \ref{eq:StochasticAction}) and its associated extremization in Eq. \ref{eq:optimumphasespace}, allow us to determine the properties of the optimal geometric phases induced by Gaussian measurements for any required boundary conditions. 
However, before exploring new subsets of geometric phases, like optimal self-closing ones, we first determine the properties of the open geometric phases under Gaussian measurements, in particular, their topological features.

For rotating null-type measurements defined in Eq. \ref{eq:kraus101} and \ref{eq:null_kraus},
preparing the qubit state along the initial measurement axis, ($\theta(0)=\Theta$, $\phi(0)=\Phi(0)$), implementing a closed loop of measurements, $\boldsymbol{n}(t)$, and post selecting the measurement readouts $r=1$ associates the path followed by the measurement with a unique trajectory on the Bloch sphere. 
This map can then be applied to the family of trajectories spanned by the initial conditions $\Theta\in [0,\pi]$.
In this way the sphere spanned by the direction of measurement operators, $\boldsymbol{n}(t)$, is mapped onto a submanifold of the Bloch sphere, via Kraus operator $M_1$.
This mapping undergoes a topological transition~\cite{gebhart2019measurement} as a function of the measurement strength $c$ from a phase in which the image of the trajectories covers the Bloch sphere, to one in which it fails to cover the entire sphere.
These two topological regimes are distinguished by a Chern number
\begin{equation}
\label{eq:ChernNumber}
C \equiv \frac{1}{2 \pi} \int_0^\pi d\Theta \int_0^1 B dt = \frac{1}{2\pi}(\chi(\pi)-\chi(0)),
\end{equation} 
with $B(\Theta, t) = \text{Im}\{ \partial_t \langle \psi |\partial_{\Theta} |\psi \rangle - \partial_{\Theta} \langle \psi | \partial_t | \psi \rangle \}$.
The Chern number can take discrete values, $C = -1$ or $C= 0$, which are directly related to the dependence of the geometric phase as shown in the last step of Eq. \ref{eq:ChernNumber}.
The topological transition of this mapping is then manifest in the measurement-induced geometric phase as a function of measurement latitude, $\chi(\Theta)$, bifurcating the function into monotonic and non-monotonic regimes. 
The mapping $[0,\pi]\ni \Theta \to \chi \in [0,2\pi]$ is in fact a mapping $S_1 \to S_1$ with winding number $w=0,1$. The latter is equivalently determined by $w=[\chi(\pi/2)-\chi(0)]/\pi \in \{0,1\}$

Gaussian measurements have been involved in researching this type of topological transition~\cite{Murch_2013}.
Dual readout Gaussian measurements on a qutrit were used to reproduce the effect of the null-type measurements on a two-level sub-system. 
Here we are concerned with Gaussian measurements in their own right, specifically in the protocol outlined in section \ref{MeasurementProtocol}. 
In this case, a continuous mapping between the two spheres is no longer constrained to the null-type post-selection and more general post-selections can be imposed on the measurement readout. 
Quite generally, we expect that a rotating measurement protocol with ($\theta_i=\Theta, \phi_i=0$) features a topological transition for post-selected record function $r(t)>0$, which ensures that the continuity of the trajectory at the initial state in the strong-measurement limit (i.e. the system is driven towards the positive eigenstate of the measured operator at all times during the measurement induced dynamics). 
For the special case when $\Theta = \frac{\pi}{2}$; the Bloch Sphere equator corresponds to an invariant subspace for every Kraus operator in the measurement sequence: states initialized on the equator remain therein. This feature of the systems accessible trajectories manifests in the available geometric phases $\{\frac{\pi}{2} n | n \in \mathbb{Z} \}$; each is associated with a definite winding number. 
This applies irrespective of the state preparation used.

\begin{figure}[h!]

\centering
\begin{subfigure}{0.5\textwidth}
    \caption{}
    \includegraphics[width=0.8\textwidth]{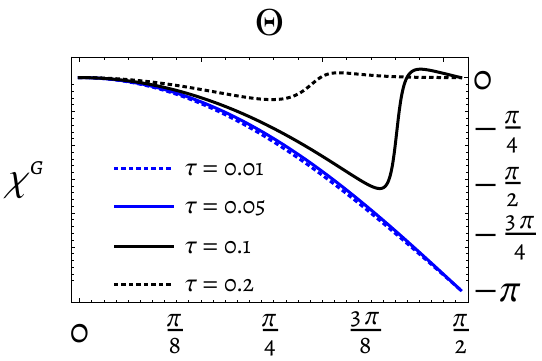}
    \label{fig:geomglobaloptimal}
\end{subfigure}
\begin{subfigure}{0.5\textwidth}
    \centering
    \caption{}
    \includegraphics[width=4.75cm]{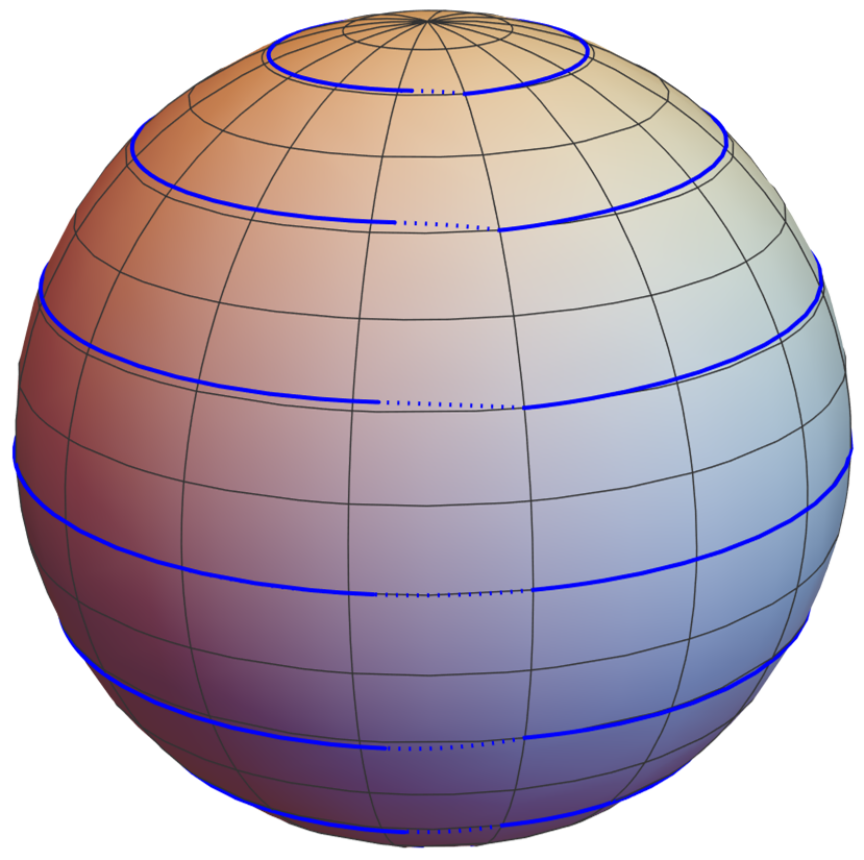}
\label{fig:geomglobalw}
\end{subfigure}
\begin{subfigure}{0.5\textwidth}
    \centering
     \caption{}
    \includegraphics[width=4.75cm]{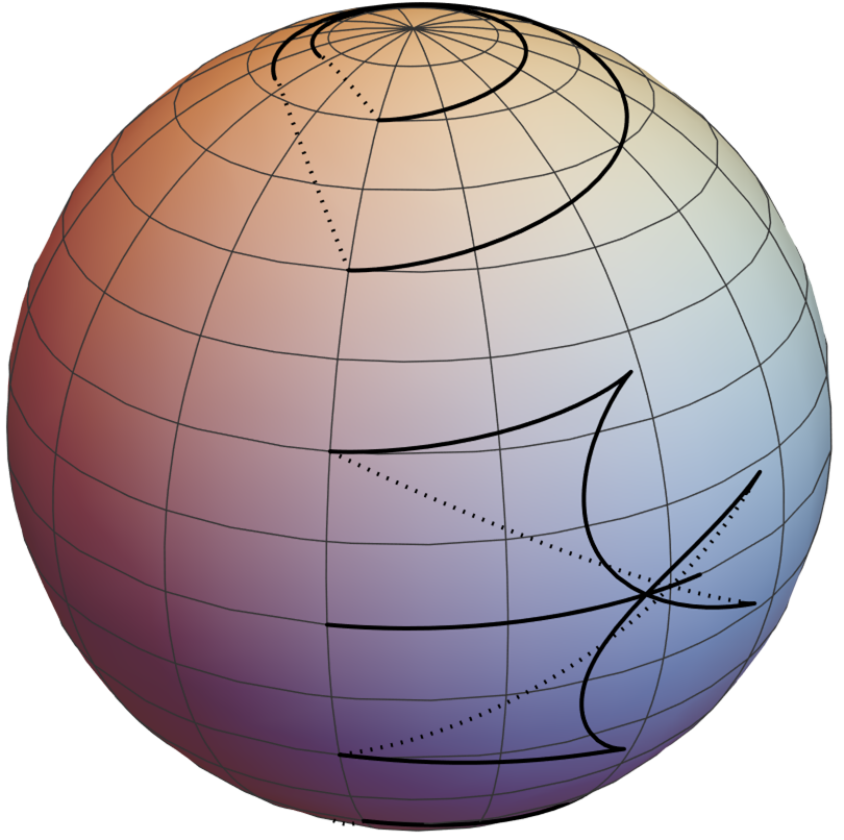}
    \label{fig:geomglobalnw}
\end{subfigure}
\caption{\textbf{Global Optimal quantum trajectories and their geometric phases}. (a) Geometric phase $\chi^g(\Theta)$  for a range of measurement strengths below (blue) and above (black) the inverse critical measurement strength $\tau_c/T \approx 0.1$.  Quantum trajectories on the Bloch sphere with closing geodesics (dotted) for $\tau/T = 0.05$ (c) and $\tau/T = 0.2$ (b). The family of trajectories covers the Bloch sphere for measurements stronger than the critical value (b) and does not otherwise (c).}
\label{fig:most-likely-open}
\end{figure}

As a first example, it is possible to ascertain the topological transition for the family of most likely trajectories spanned by initial states that coincide with the measurement axis. 
These most-probable post-selected optimal trajectories are obtained from the observation that at each time-step, the most probable outcome is given by $\langle r \rangle = a(\Theta, t)$ (cf. Eq. \ref{eq:conditionalprob}). Hence, substituting the time continuous version, $ r(t) = a(t)$ from Eq. \ref{eq:meanreadout} into Eq. \ref{eq:SMEspherical1}-\ref{eq:SMEspherical3}, with the initial conditions $\theta(0) = \Theta$ and $\phi(0) = \Phi(0)$, produces the required optimum quantum trajectory. 
A topological transition is observed in this case, as illustrated in Fig. \ref{fig:most-likely-open}, where the family of quantum trajectories --- including the closing geodesic--- for strong measurements (small $\tau/T$) wrap the Bloch sphere (Fig. \ref{fig:geomglobalw}, while for weak measurements (large $\tau/T$) they do not (Fig. \ref{fig:geomglobalnw}. 
The transition measurement strength is determined numerically from the dependence $\chi^g(\Theta)$, which is reported in Fig \ref{fig:geomglobaloptimal}. 
For $\tau < \tau_c$ the geometric phase evolve continuously from $0$ to $-\pi$, while for $\tau>\tau_c$, $\chi(\pi/2)=\chi(0)$. We estimate that the transition occurs for $\tau_c/T \approx 0.1$, 
The transition is further confirmed by a direct numerical evaluation of the Chern number using Eq. \ref{eq:ChernNumber}.

\begin{figure}
\begin{subfigure}{0.5 \textwidth}
\centering
\caption{}
\includegraphics[width=6cm]{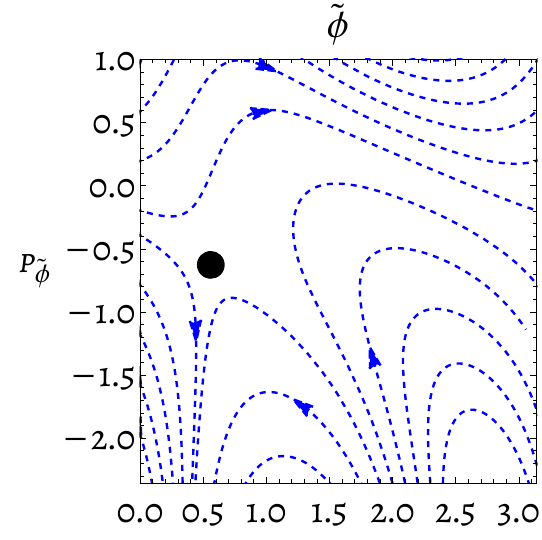}
\label{fig:phasespace01}
\end{subfigure}
\begin{subfigure}{0.5 \textwidth}
\centering
\caption{}
\includegraphics[width=6cm]{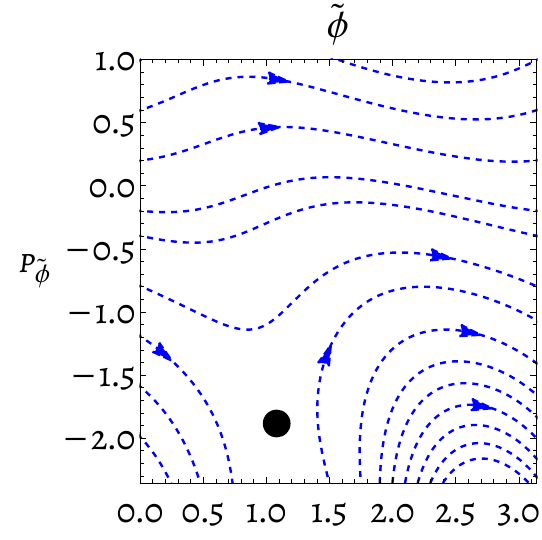}
\label{fig:phasespace03}
\end{subfigure}
\begin{subfigure}{0.5 \textwidth}
\centering
\caption{}
\includegraphics[width=4.75cm]{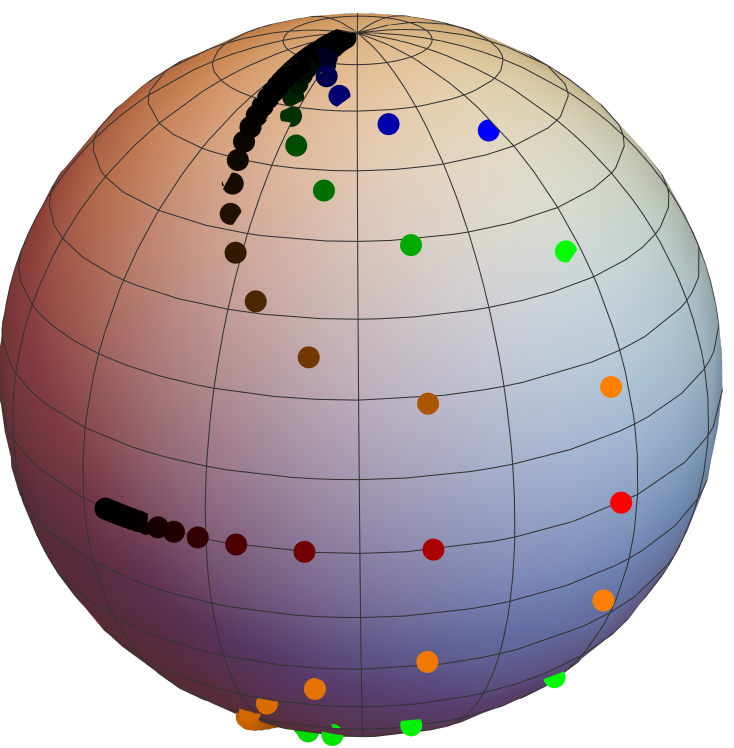}
\label{fig:eqpoints}
\end{subfigure}
\caption{\textbf{Equilibrium state}. Flow of Hamilton's equations in the CDJ Phase Space at $\theta=\pi/2$ (Eq. \ref{eq:optimumphasespace}) in the co-rotating coordinate system for $\tau = 0.1$, $\Theta = \frac{\pi}{2}$ (panel a) and $\tau = 0.3$, $\Theta = \frac{\pi}{2}$ (panel b).  The equilibrium point is indicated by the black dot. Panel c): Dependence of the equilibrium points ($\theta_e, \phi_e$) on  $\Theta$ and $\tau$. For each value of  $\Theta$ ($ \frac{\pi}{6}$ (blue),$\frac{2 \pi}{7}$ (green),$\frac{2\pi}{5} $ (orange),$\frac{\pi}{2} $(red), $\frac{3 \pi}{5} $(orange),$\frac{5 \pi}{7}$ (green)) darker shades corresponding to weaker measurements from $\tau/T= 0.1 $ to $\tau/T= 10$.}
\label{fig:phasespace}
\end{figure} 

A second kind of variation of the protocol concerns the state preparation.
From numerical simulation of a range of cases, it appears that the initialization of the system does not affect the topological nature of the transition and the associated phenomenology of the geometric phase provided the state initialization $\theta_i(\Theta(0))$ spans the entire range of the polar angle, is monotonic, and satisfies $\theta_i(0) = 0$, $\theta_i(\frac{\pi}{2})=\frac{\pi}{2}$, and $\theta(\pi) = \pi$ with $\lim_{\tau \to 0} \phi_i = \Phi(0)$.
Note we are explicitly allowing for the possibility of a measurement strength-dependent state preparation. 
We are particularly interested in using this freedom to choose a new state preparation that, similar to the choice $\theta_i = \Theta$ $\phi_i = \Phi(0)$, will also continuously recover the projective measurement limit - where states are initialized along the measurement axis and meticulously follow the axis for their entire evolution. 
This then requires a state initialization that gives $\theta_i \to \Theta(0)$ and $\phi_i \to \Phi(0)$ in the strong measurement limit.

\begin{figure}[h!]
\begin{subfigure}{0.45 \textwidth}
\centering
\caption{}
\includegraphics[width=0.9\textwidth]{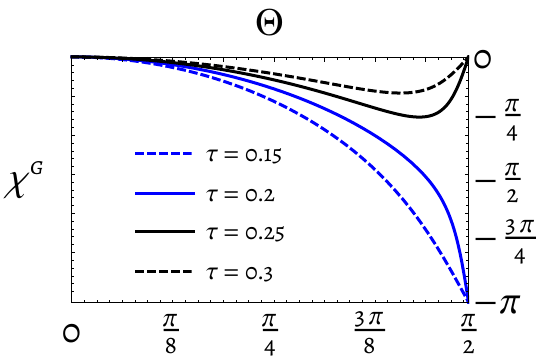}
\label{fig:geomeqopen}
\end{subfigure}
\begin{subfigure}{0.45 \textwidth}
\centering
\caption{}
\includegraphics[width=0.5\textwidth]{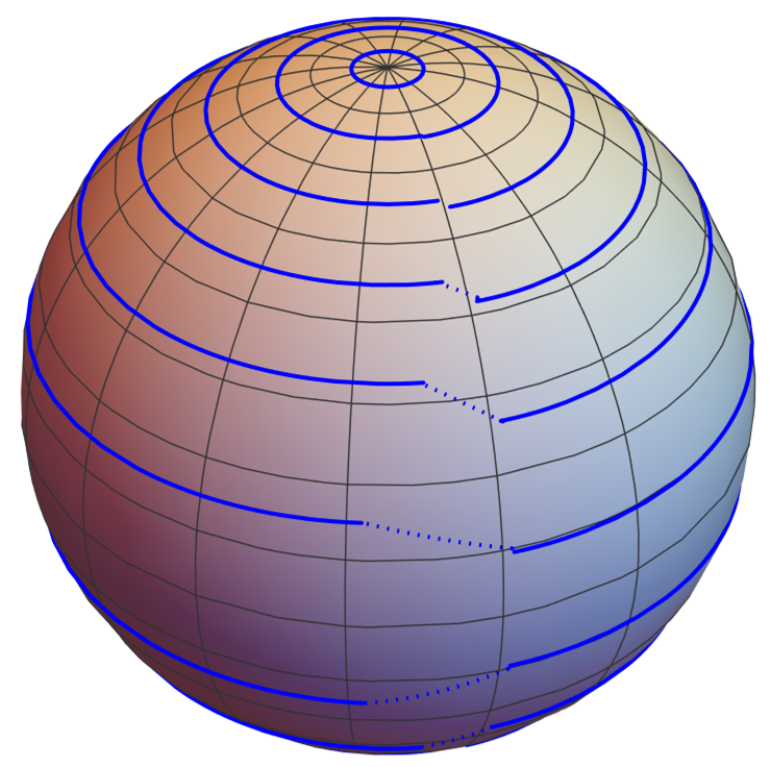}
\label{fig:geomglobalw}
\end{subfigure}
\begin{subfigure}{0.45 \textwidth}
\centering
\caption{}
\includegraphics[width=0.5\textwidth]{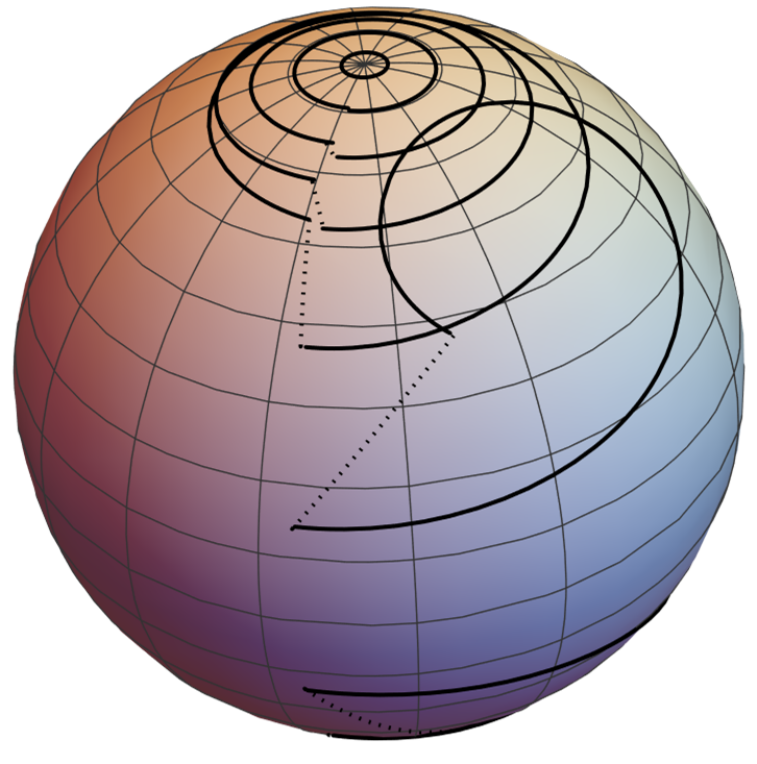}
\label{fig:geomglobalnw}
\end{subfigure}
\caption{\textbf{Quantum trajectories and geometric phases with equilibrium state initialization}. Panel a) Geometric phase $\chi(\Theta)$ for a range of measurement strengths below (blue) and above (black) the inverse critical measurement strength $\tau_c \approx 0.22$. Quantum trajectories on the Bloch Sphere with closing geodesics (dotted) for $\tau = 0.15$  (b)  $\tau = 0.3$ (c). The family of trajectories covers the Bloch sphere for measurements stronger than the
critical value (b) and does not otherwise (c).}
\label{fig:EquilibriumOpenTransition}
\end{figure}

A natural case, which will be relevant later on for closed geometric phases,  has the initial state chosen to coincide with a fixed point (so $\theta_i = \theta_e$ and $\phi_i = \phi_e$) of the optimal trajectory dynamics in the co-rotating coordinate frame. 
This equilibrium point in the co-rotating dynamics ($\theta_e,\phi_e$) is defined by $\dot{\tilde{\phi}}=0$, $\dot{\tilde{\theta}}=0$, $\dot{\tilde{p}}_\phi=0$, $\dot{\tilde{p}}_\theta=0$.
Hamilton's Equations for the action (Eq. \ref{eq:rotating action}) can be solved to determine a closed-form expression,
\begin{align}\label{eq:eqpoint}
     \theta_e &= \tan ^{-1}\left(\frac{\tan (\Theta)}{\sqrt{4 \pi ^2 \tau ^2+1}}\right), \\
    \phi_e &= - \tan^{-1}(2 \pi \tau).
\end{align}
Fig. \ref{fig:phasespace}, which reports the phase space flow diagram from Hamilton's equations for the action in Eq. \ref{eq:rotating action} for different values of the measurement strength [panels (a) and (b)] at $\theta=\Theta_0=\pi/2$. 
For increasingly strong measurements $\phi_e$ tends toward the measurement axis. 
The position of equilibrium points for generic latitudes on the Bloch sphere is reported in Fig. \ref{fig:eqpoints}, showing that this same limiting behaviour continues across the entire Bloch Sphere.
The quantum trajectories for this modified protocol (with a generic choice of post-selection $r(t)=1$) are reported in Fig. (\ref{fig:EquilibriumOpenTransition}[b-c]), where small $\tau/T$  (panel b) led to trajectories wrapping the Bloch sphere and large  $\tau/T$ (panel c) don't. 
Similarly to the case reported in Fig. \ref{fig:most-likely-open}, we can identify a topological transition from the behaviour of $\chi$ as a function of $\Theta$ [cf. Fig. \ref{fig:geomeqopen}], which gives a critical measurement strength $\tau_c/T\approx 0.22$.
From these examples, it emerges that, despite variations in the exact value of $\tau_c$, the fundamental characteristics of the transition are unchanged for a wide range of state preparation protocols.

\section{Topological Transition in the optimal closed geometric phase}\label{sec:OptimalTransition}

We now use the developed action formalism to investigate the topological features of the closed geometric phase. 
This involves considering the set of self-closing trajectories generated solely by continuous monitoring, without a projective measurement step.
Since self-closing trajectories are not generally achievable by conditioning on any single measurement record function, we post-selected most likely self-closing trajectory, corresponding to the most-probable closed geometric phase attained during measurement. 
We denote the optimum phase as $\chi^{\text{opt}}$. 
On the Bloch Sphere equator, the behaviour of $\chi^{\text{opt}}$ is fundamentally tied to the topology of $S^{1}$.
Attaining only two possible values corresponding to the winding number of the associated quantum trajectory, either $-\pi$ or $0$. 

Care must be taken to find a state preparation that will produce $\chi^{\text{opt}}$ that recovers the projective measurement limit, 
\begin{eqnarray} \label{eq:geomlimits}
    \lim_{\tau \to 0 } \chi^{\text{opt}} \approx -2 \pi  \sin ^2\left(\frac{1}{2} \Theta \right), \lim_{\tau \to \infty} \chi^{\text{opt}} \approx 0.
\end{eqnarray}
The state preparation $\theta(0) = \Theta$ and $\phi(0)=\Phi$, applied to Eq. \ref{eq:optimumphasespace}, with self-closing boundary conditions produce a variety of candidate optimal geometric phases that satisfy Eq. \ref{eq:geomlimits}. 
In the regime of strong measurements, one might anticipate an increased likelihood of candidate solutions with  $\lim_{\tau \to 0 } \chi^{\text{opt}} \approx -2 \pi  \sin ^2\left(\frac{1}{2} \Theta \right)$.
However, numerical assessment of the stochastic action reveals that these solutions remain vanishing improbable even in the strong measurement limit.
This phenomenon can be understood upon examining the characteristics of these candidate optima:
spending the majority of their lifetime (approximately $T$) following the equilibrium trajectory (Eq. \ref{eq:eqpoint}), punctuated with rapid transitions to and from the measurement axis at the beginning and end of the protocol.
It is these rapid transitions that suppress the likelihood of the winding solution. 
Consequently, solutions associated with a vanishing winding number dominate. 
This scenario leads to a $\chi^{\text{opt}}$ that does not satisfy the prescribed condition.
Instead, a suitable state preparation is provided by the equilibrium point introduced in Eq. \ref{eq:eqpoint}.
The equilibrium points span the entire range of latitudes, with $\theta \in [0, \pi]$.
As we shall later demonstrate, with this particular initialization, the value of $\chi^{\text{opt}}$ adheres to the required limits given in Eq. \ref{eq:geomlimits}.
This establishes the mapping between measurement parameters  $(\Theta, \tau)$ and the optimal self closing quantum trajectory and it's associated closed geometric phase $\chi^{\text{opt}}$ that we use to establish a new topological transition.

\subsection{Topological Transition}

As discussed for the case of the open geometric phase, the topological properties of the mapping of $\boldsymbol{n}(t) \to q(t)$ is dictated by the fixed points at $\Theta=0$ and $\Theta=\pi$. When $\Theta = \frac{\pi}{2}$, the simplest topological features can be investigated since here the accessible trajectories are each associated with a definite winding number $n$ indexing the available closed geometric phases $\{\frac{\pi}{2} n \ | \  n \in \mathbb{Z} \}$.
Hamilton's equations for the Stochastic Action, Eq. \ref{eq:optimumphasespace}, restricted to the Bloch Sphere equator, have multiple solutions after imposing boundary conditions  $\phi(0) = \phi_e$ and $\phi(T) = \phi_e + 2 \pi n $, generating a set of candidate most-likely self-closing quantum trajectories and phases. 
A single candidate solution is generated for each value of $n$ corresponding to a local minimum of the action (or equivalently a local maximum of the probability density, $\mathfrak{p}$ over the set of quantum trajectories).

\begin{figure}[h!]
\begin{subfigure}[t]{.5\textwidth}
\caption{}
\includegraphics[width=8.5cm]{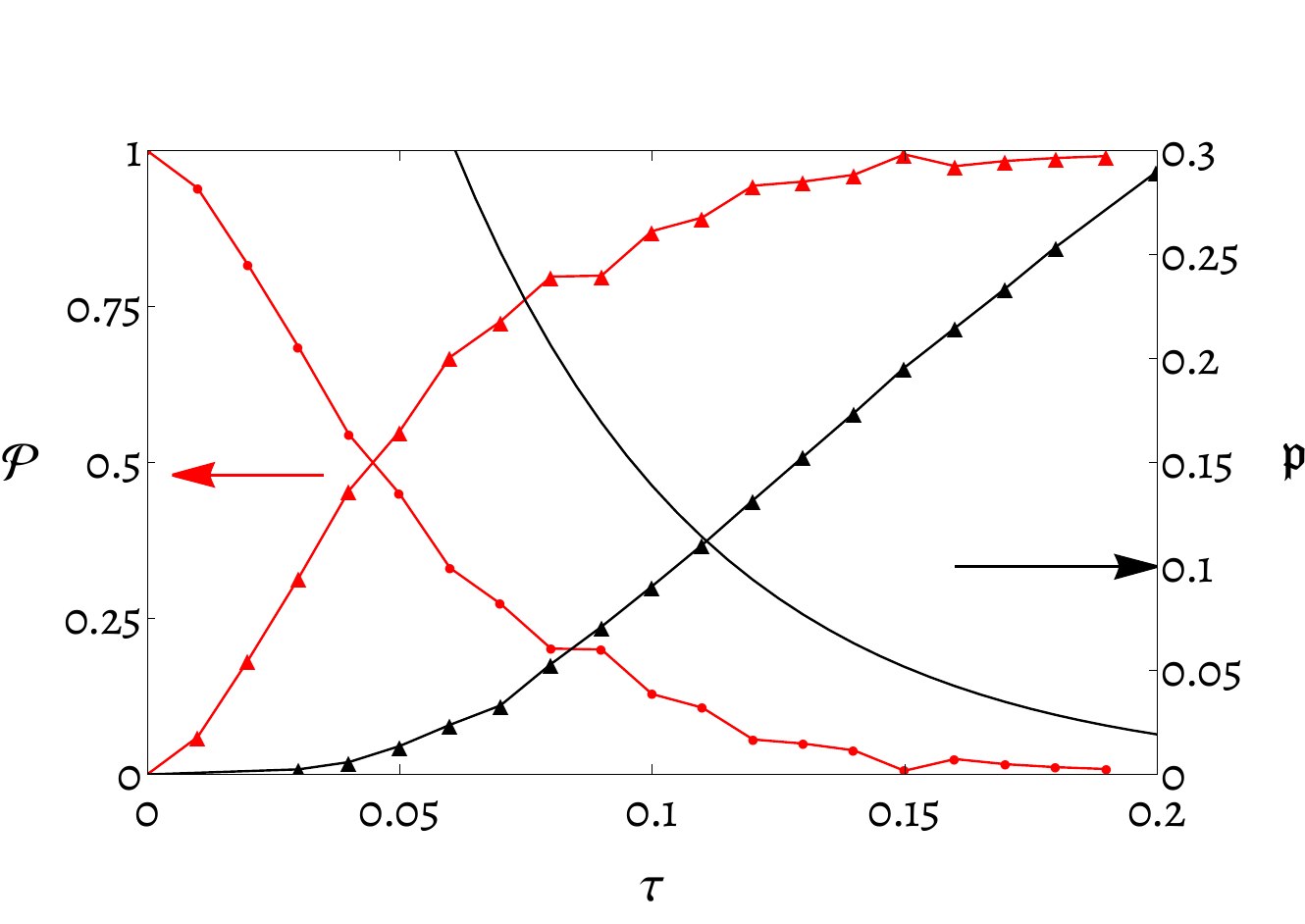}
\label{fig:crossoverplot}
\end{subfigure}
\begin{subfigure}[t]{.5\textwidth}
\caption{}
\includegraphics[width=.9\textwidth]{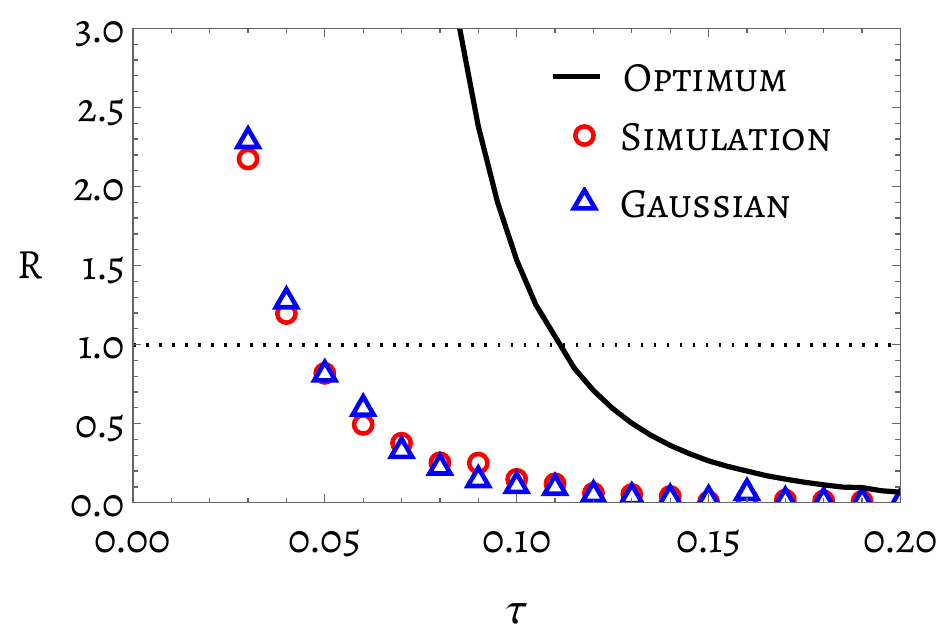}
\label{fig:ratioplot}
\end{subfigure}
\caption{\textbf{Stochastic properties of winding and non-winding quantum trajectories on the Bloch Sphere Equator.} Panel a)  Probability measure's $\mathfrak{p^{n=1}}$(solid black) and $\mathfrak{p^{n=0}}$ (black triangles) as functions of $\tau$. Normalised self-closing conditional probability $\mathcal{P}$ (state initialization ($\theta_i = \theta_e,\phi_i = \phi_e)$) for non-winding trajectories (red triangles) and winding trajectories (red). $\mathcal{P}$ is obtained from numerical simulation of Eq. \ref{eq:kraus101} and \ref{eq:gauss_simple} with 100 time-steps and 500 quantum trajectories with bin size $\Delta \phi = 0.1$. Panel b) The $\tau$ dependence of the ratios $\mathfrak{R}$ and $R$ (c.f. section $\ref{sec:gaussian corrections}$) calculated using numerical data in panel a and the results of Gaussian corrections of section \ref{sec:gaussian corrections}.}
\label{fig:ratio}
\end{figure}

To determine which candidate solution occurs with a higher probability, we evaluate and compare their actions.
The solution corresponding to  $n=1$, equivalent to the equilibrium quantum trajectory in co-rotating coordinates, is $ \phi(t)^{n = 1}_{\Theta = \frac{\pi}{2}} = 2 \pi t - \arctan{2 \pi \tau}$.
By substituting this solution into Eq. \ref{eq:StochasticAction}, we find the associated probability density is given by $\mathfrak{p}^{n = 1}_{\Theta = \frac{\pi}{2}} = e^{-2 \pi ^2 \tau }.$ 
For $n= 0$, a numerical solution to Eq. \ref{eq:optimumphasespace} is used to find $\phi(t)^{n = 0}_{\Theta = \frac{\pi}{2}}$, which is then substituted directly into the stochastic action to evaluate $\mathfrak{p}^{n =0}_{\Theta = \frac{\pi}{2}}$. 
Both of these probability densities are plotted in Fig. 
\ref{fig:crossoverplot}.
Solutions for other values of $n$ are found to have strictly lower probabilities.
We, therefore, focus our analysis on the competition between the candidate optimum trajectories indexed by $n=0$ and $n=1$.
The value of $\chi^{\text{opt}}$ is determined by the competition between the two most prominent candidate optimums.
As evidenced by Fig. \ref{fig:crossoverplot}, there is a measurement strength, $\tau_c/T \approx 0.11$, at which the optimum geometric phase jumps discontinuously from $0$ to $-\pi$. 

\begin{figure}\label{fig:TopTransitionClosed}
\begin{subfigure}[b]{0.5\textwidth}
\centering
\caption{}
\includegraphics[width=4.75cm]{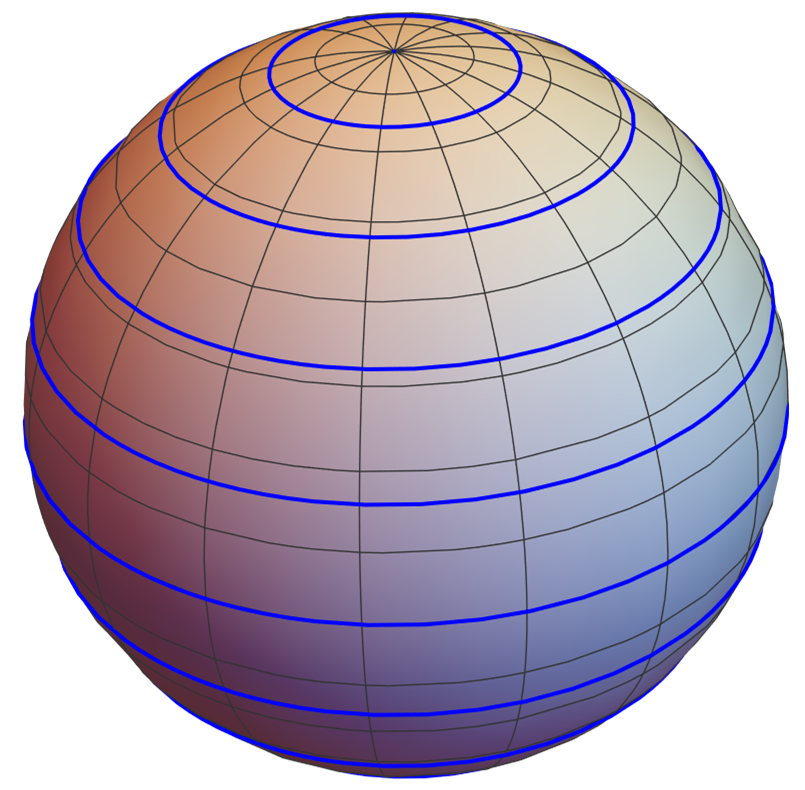}

\label{fig:WindingFamily}
\end{subfigure}
\begin{subfigure}[b]{0.5\textwidth}
\centering
\caption{}
\includegraphics[width =4.75cm]{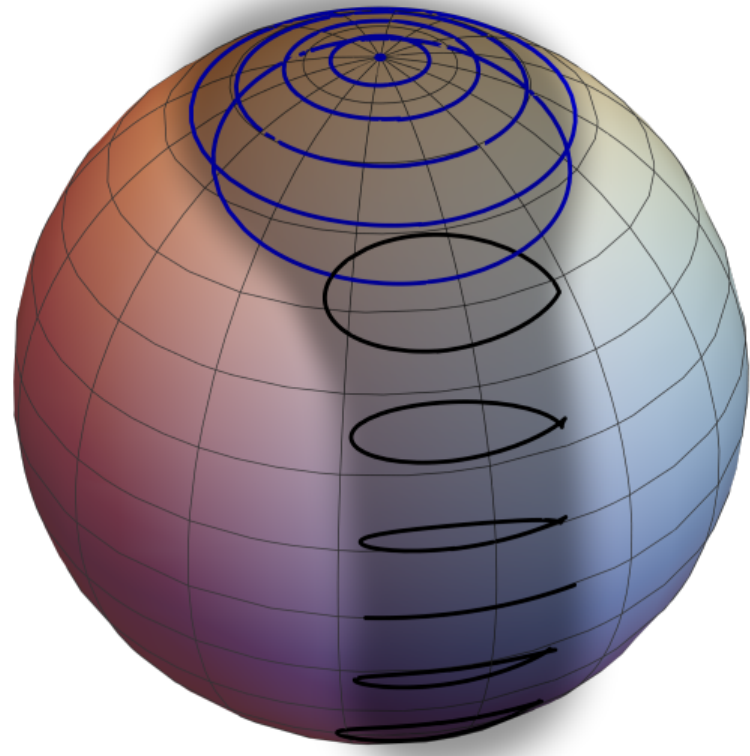}

\label{fig:NonWindingFamily}
\end{subfigure}
\begin{subfigure}[b]{0.5\textwidth}
\centering
\caption{}
\includegraphics[width=9cm]{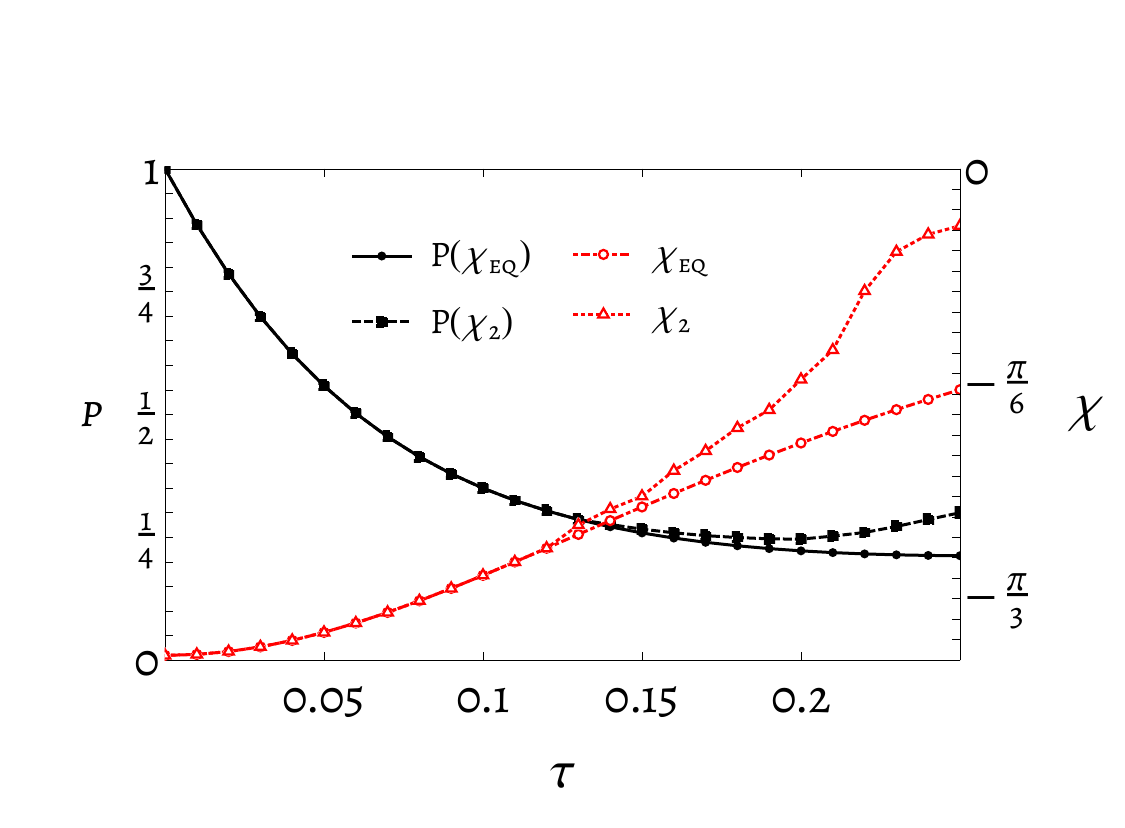}

\label{fig:merging}
\end{subfigure}

\caption{\textbf{ Optimal self-closing quantum trajectories.}  Equilibrium quantum trajectories  $\phi_{eq}$ [panel (a)] and $\phi(t)^{n = 0}$ [panel (b)]for various choices of $\Theta$, with $\tau/T = 0.2$. The shaded region in panel (b) is a guide to the eyes highlighting the $C = 0$ submanifold on the Bloch sphere. The trajectories with $C=1$ cover the whole sphere as shown in panel (a). Panel c) Candidate optimum geometric phases, $\chi_{eq}$ and $\chi_{2}$ and corresponding probabilities, $P(\chi_{eq})$ and the competing optimum $P(\chi_{2})$, at $\Theta = 0.9 < \Theta_c$ as a function of $\tau$. The two candidate solutions merge into a single q-trajectory before the value of $\tau_c$.}
\end{figure}

We now investigate if a transition of topological number in the optimal geometric phase occurs across the whole Bloch sphere.
To address this question, we study the subset of solutions for Eq. \ref{eq:optimumphasespace} that generate optimal self-closing quantum trajectories (with $\theta(0)= \theta_e$, $\phi(0) = \phi_e $ and $\chi(0)=0$) for all measurement latitudes.
We identify one family of solutions ($\phi_{eq}$) smoothly connected to $\phi(t)^{n = 1}_{\Theta = \frac{\pi}{2}}$, specified by $\phi_{eq} = \phi_e + 2\pi t$ and $\theta(t) = \theta_e$.
This closed-form solution may be substituted back into Eq. \ref{eq:optimumphasespace} to find the corresponding geometric phase,\begin{equation}
       \chi^{n=1} = -2 \pi  \sin ^2\left(\frac{1}{2} \tan ^{-1}\left(\frac{\tan (\Theta)}{\sqrt{4 \pi ^2 \tau ^2+1}}\right)\right),
\end{equation} which is proportional to the solid angle of the spherical cap defined at the latitude $\theta_e$. 
We note that $ \lim_{\tau \to 0}\chi^{n=1} =  -2 \pi  \sin ^2\left(\frac{1}{2} \Theta \right)$, and that $\chi^{n=1}$ decreases monotonically from 0 to $-\pi$ with increasing $\Theta$. 
Similarly, we evaluate $S$ on this family of solutions, finding the probability density \begin{equation}
   \label{eq:phase1} \mathfrak{p}^{n=1} = \exp \left(-\frac{2 \pi ^2 \tau  \sin ^2(\Theta)}{2 \pi ^2 \tau ^2 \cos (2 \Theta)+2 \pi ^2 \tau ^2+1}\right).
\end{equation}
Note that $\mathfrak{p}^{n=1} \to 1$ for $\tau \to 0$.  These sets of solutions for different $\Theta$ form a sub-manifold of the Bloch sphere with $C=1$, independent of the measurement strength (see Fig. \ref{fig:WindingFamily}).
We may identify a second family of quantum trajectories: the most likely solutions excluding $\phi_{eq}$\footnote{As for the solutions at the equator, multiple solutions of the Euler-Lagrange equations exist for given boundary conditions, corresponding to local minima of the actions} (see Fig. \ref{fig:NonWindingFamily}). 
This set of solutions, $\phi(t)^{n = 0}$, is determined by numerically computing the stochastic action in Eq. \ref{eq:StochasticAction} and has no closed-form expression for either the accrued geometric phase or the associated probability density. 
The mapping between $(\Theta(t), \Phi(t))$ and $(\theta(t), \phi(t))$ in this case is always characterized by $C = 0$. The optimal (most likely) value of the geometric phase, $\chi^{\text{opt}}$, is then determined by the competition between these two families of solutions.

\begin{figure}[h!]
\centering
\includegraphics[width=0.45\textwidth]{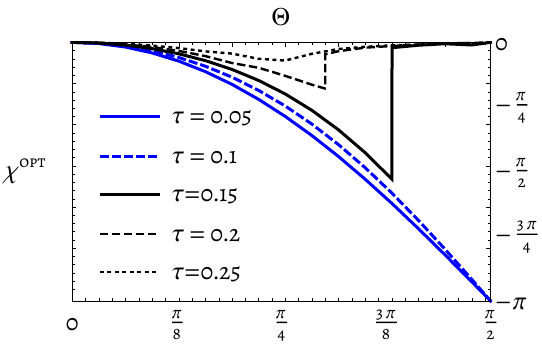}
\caption{\textbf{Optimal geometric phases, $\chi^{\text{opt}}$, as a function of $\Theta$ for a range of measurement strengths.} The critical measurement strength $\tau_c$ distinguishes the behaviour $\chi^{\text{opt}}(\Theta=\pi/2)=0$ and $\chi^{\text{opt}}(\Theta=\pi/2)=-\pi$. for $\Theta>\Theta_C$, the geometric phase can exhibit a jump for $\Theta>\Theta_C$.}
\label{fig:optGeom}
\end{figure}

For strong measurements ($\tau \ll T$), solutions $\phi(t)^{n = 0}$ are less probable than $\phi_{eq}$, so the optimal geometric phase is given by Eq. \ref{eq:phase1}. 
In the weak measurement regime ($\tau \gg T$), we have the converse, solutions $\phi(t)^{n = 0}$ are more likely.
We define a $\tau$ dependent function, $\Theta_{\textrm{jump}}(\tau)$, that separates two types of behaviour in $\chi^{\text{opt}}$ as it approaches the strong measurement limit. 
At $\Theta_{\textrm{jump}}(\tau)$, $\chi^{\text{opt}}$ jumps discontinuously to the value determined by $\phi_{eq}$, with $\tau_c$ marking the smallest value of $\tau$ for which this discontinuous jump occurs.
This jump in the geometric phase is shown in Fig. \ref{fig:optGeom}.
We name the largest value of $\Theta_{\textrm{jump}}$, $\Theta_C$, and it is determined to be $\Theta_C \approx 0.95$.

Below $\Theta_{C}$, $\chi^{\text{opt}}$ tends smoothly towards the geometric phase associated with $\phi_{eq}$. 
Numerical evidence suggests that the family of quantum trajectories $\phi(t)^{n = 0}$ merge smoothly with the associated trajectory $\phi_{eq}$.
This behaviour is illustrated in Fig. \ref{fig:NonWindingFamily}, where the blue-coloured quantum trajectories denote the region in which the family of trajectories $\phi(t)^{n = 0}$ merges into $\phi_{eq}$. 
This behaviour is quantified in Fig. \ref{fig:merging}, which shows how, at $\Theta <\Theta_C$, the geometric phase from the $\phi(t)^{n = 0}$ family of trajectories coincides with that from $\phi_{eq}$ at $\tau<\tau_c$.
Crucially, this merger occurs above the critical measurement strength $\tau_c$, suggesting that the value of $\tau_c$ as determined on the Bloch sphere equator does correspond to a topological transition across the entire Bloch Sphere. 
This transition is manifest in the behaviour of $\chi^{\text{opt}}$ as a function of $\Theta$ (see Fig. \ref{fig:optGeom}), where $\tau_c$ separates phases that are continuous and monotonically decreasing (from 0 to $\pi$) and those which are non-monotonic (0 to 0).
The transition in the topological number for optimal self-closing trajectories and the corresponding discontinuity in the geometric phase are distinct from the open geometric phase transition. 
While the latter is associated with a vanishing post-selection probability at the critical point~\cite{gebhart2019measurement}, for self-closing trajectories, two trajectories from distinct families become equally likely at the transition point. 

This topological transition is well defined in terms of the most likely trajectories belonging to either of the two families identified above. However, any experiment would not be able to access the most likely trajectories directly. The set of trajectories to be compared must necessarily include an ensemble of trajectories that are equivalent, up to the precision of the experiment. 
The averaged geometric phase is expected to display a crossover as opposed to a sharp transition at $\tau_c$. In a scenario with finite experimental precision, the value of $\tau_c$ will therefore be smeared out.
However, a transition can still be identified in terms of which of the ensembles (each associated with a distinct optimal geometric phase) is most probable.
We calculate the effect this has on the value of $\tau_c$ in the next section.

\section{Gaussian Corrections}\label{sec:gaussian corrections}

To incorporate multiple trajectories in the picture we include Gaussian corrections in the analyses of optimal self-closing trajectories.
This method allows us to account for the effect of solutions that deviate slightly from the optimal solutions while still satisfying the required boundary conditions (and hence can be associated with distinct geometric phases). Operationally, the identification of  $\tau_c$ when including extra trajectories is achieved by replacing a direct comparison of the measure of two individual quantum trajectories,  with a comparison of the state transition probabilities,
\begin{equation}
    \mathfrak{R} = \frac{\mathfrak{p}(\phi^{n=1}_{\Theta = \frac{\pi}{2}})}{{\mathfrak{p}(\phi^{n=0}_{\Theta = \frac{\pi}{2}})}} \to R=\frac{\mathcal{P}^{\text{eq}}(2\pi +\phi_e|\phi_e)}{\mathcal{P}^{n=0}(\phi_e|\phi_e)},
\end{equation} 
so that $\mathcal{P}^{eq}(2\pi +\phi_e|\phi_e)$ is the self-closing transition probability  evaluated for trajectories close to $\phi_{eq}$ with geometric phases $-\pi$. $\mathcal{P}^{n=0}(\phi_e|\phi_e)$ is the self-closing transition probability evaluated for trajectories close to $\phi^{n=0}$ which are associated with a vanishing geometric phase. Here, we limit our analysis of Gaussian corrections to states initialized with $\Theta=\pi/2$ since the transition and its critical value are determined by the equatorial dynamics. The dynamics are then fully constrained to one dimension, parametrized by $\phi$.

$\mathcal{P}^{\text{eq}}$ and $\mathcal{P}^{n=0}$ can be evaluated using a saddle point approximation around their associated candidate optimum quantum trajectory. 
Calculating a saddle point approximation is a standard procedure~\cite{altland_simons_2010} that consists in expanding the action around a chosen optimum solution by rewriting $\bold{q} = \bold{q}^{*} + \delta \bold{q}$ and $\bold{p} = \bold{p}^{*} + \delta \bold{p}$, where $\delta \bold{q}$ and $\delta \bold{p}$ are small deviations from the optimal values $\bold{q}^{*}$ and $\bold{p}^{*}$. The Gaussian terms in the path integral may then be evaluated, neglecting higher-order corrections. The result is an approximated probability, \begin{equation}
\mathcal{P} = \mathcal{N} \int \mathcal{D}\textbf{q} \mathcal{D} \textbf{p} \ e^{S[\textbf{q},\textbf{p}]} \approx \mathcal{N}
\sqrt{\frac{(2\pi)^n}{\det(\bold{A}|_{\bold{q}^{*},\bold{p}^{*}} )}} e^{S[\bold{q}^{*},\bold{p}^{*}],}
\end{equation}
where $\bold{A}|_{\bold{q}^{*},\bold{p}^{*}}$ is the functional hessian of the CDJ action evaluated on the optimum path with zero Dirichlet boundary conditions and $\mathcal{N}$ is an overall normalisation factor. 

Our analysis is simplified by expressing the stochastic path integral in the Lagrangian formulation (Eq. \ref{eq:path}), despite the presence of the state-dependent functional measure in Eq. \ref{eq:probabilitydensityphi}. Simplifying Eq. \ref{eq:path}, \ref{eq:probabilitydensityphi} and \ref{eq:lagrangian} with $\theta(t) = \frac{\pi}{2}$ and $\Theta = \frac{\pi}{2}$, and recalling the value of $\theta_e$, the path integral we required for $\mathcal{P}^{\text{eq}}(2 \pi +\phi_e| \phi_e)$ is,
\begin{align}
&\mathcal{P}(2 \pi +\phi_e| \phi_e) \propto   \\ &\int \mathcal{D} \phi \sqrt{\frac{\tau}{2 \pi \sin^2(2 \pi t - \phi)}} e^{\int \big(-\frac{1}{2} \tau  \dot{\phi}^2 \csc ^2(2 \pi  t-\phi)+\dot{\phi} \cot( 2 \pi  t-\phi)\big) dt} \\ \nonumber  &\approx  e^{S[\phi_{eq}]}\int\mathcal{D} \delta \phi \sqrt{\frac{\tau}{2 \pi \sin^2(2 \pi t - \phi_{eq})}}e^{\int \delta \phi \bold{A}|_{\phi_{eq}} \delta \phi  dt}.
\end{align}
Following the method described in~\cite{PhysRevE.99.032125}, we apply a time reparameterization around the equilibrium trajectory, so the new time variable $u$ is determined by $u(t) = \frac{1}{\tau}\int_0^t \sin(2\pi t' - \phi_{eq}(t')) dt'$ with $u(0)= 0$. This acts as a local-scale transformation that eliminates the state-dependent functional measure, simplifying the path integral to,
\begin{eqnarray}
\mathcal{P}(2 \pi + \phi_e| \phi_e) \approx \mathcal{N}e^{S[\phi_{eq}]}\int\mathcal{D} \delta \phi e^{\int \delta \phi \bold{\Sigma}|_{\phi_{eq}} \delta \phi  du}.
\end{eqnarray}
This may then be evaluated as,
\begin{align}\label{eq:GCfactors}
e^{S[\phi_{eq}]}&\int\mathcal{D} \delta \phi e^{\int \delta \phi \bold{\Sigma}|_{\phi_{eq}} \delta \phi  du} = e^{S[\phi_{eq}]} \Big( \det \bold{\Sigma}|_{\phi_{eq}}\Big)^{-\frac{1}{2}} \\ \nonumber &= e^{S[\phi_{eq}]}\Big( \frac{\det \big[ \Sigma|_{\phi_{eq}} \big]}{\det \big[ \frac{d^2}{du^2} \big]}\Big)^{-\frac{1}{2}} \Big( \det_{\zeta} \big[ \frac{d^2}{du^2} \big] \Big)^{-\frac{1}{2}},
\end{align}
where the final functional determinant is Riemann-Zeta regularized~\cite{altland_simons_2010}. This regularized determinant would normally be absorbed in the functional measure, however, given the time parametrization we employed depends on the value of the state parameter we must include the state-dependent part of the Riemann Zeta regularised contribution of this term which can be expressed as~\cite{PhysRevE.99.032125} 
\begin{equation}
    \text{det}_{\zeta} \frac{d^2}{du^2} = \prod \pi^2 n^2 \prod \frac{1}{u(T)^2} = |u(T)|=\frac{4 \pi ^2 \tau }{4 \pi ^2 \tau ^2+1}.
\end{equation}
The ratio of functional determinants may be calculated by the Gelfand-Yaglom method~\cite{Kirsten_2004} giving,
\begin{equation}
    \frac{\det \big[ \Sigma|_{\phi_{eq}} \big]}{\det \big[ \frac{d^2}{du^2} \big]} = \frac{f(u|{t=1})}{f^{0}(u|{t=1})} = \frac{\tau  \sinh \left(\frac{\sqrt{4 \pi ^2 \tau ^2+1}}{\tau }\right)}{\sqrt{4 \pi ^2 \tau ^2+1}},
\end{equation}
where $\Sigma|_{\phi_{eq}} f(u) = \lambda f(u) \text{ and }
    \frac{d^2}{du^2} f^0(u) = \lambda^0_i f^0(u) $
with initial conditions $f^{(0)}(0)=0$ and $\dot{f^{(0)}}(0)=1$. 
For the winding trajectory, $\phi_{eq} = 2 \pi t - \arctan(2 \pi \tau)$, we find a closed form expression for the probability density,
\begin{equation}
 P(2 \pi + \phi_e| \phi_e) \propto e^{-2 \pi ^2 \tau}  \left( \frac{ 4 \pi ^2 \tau^2 \sinh \left(\frac{\sqrt{4 \pi ^2 \tau ^2+1}}{\tau }\right)}{(4 \pi ^2 \tau ^2+1)^{\frac{3}{2}}}\right)^{-\frac{1}{2}} .
\end{equation}

For $\mathcal{P}^{n=0}(\phi_e|\phi_e)$, we employ the same method, however, finding no closed form solution for the saddle point, we resort to a numerical approximation of each of the three contributing factors (equivalent to Eq. \ref{eq:GCfactors} evaluated around $\phi^{n=0}$) directly evaluating $e^{S[\phi^{n=0}]}$ and $|u(T)|$ numerically while approximating the ratio of the two functional determinants using the smallest $N$ eigenvalues,
\begin{equation}
   \frac{\det \big[ \Sigma|_{\phi^{n=0}} \big]}{\det \big[ \frac{d^2}{du^2} \big]} \approx \frac{\prod_i^{N} \lambda_i}{\prod_i^{N} \lambda^{0}_{i}}.
\end{equation}
Using these results, the ratio $R$ is plotted in Fig. \ref{fig:ratioplot}, where $R$ is compared to $\mathfrak{R}$ and to the value of $R$ computed using numerical simulations. we find that the effective value of $\tau^{\text{eff}}_c \approx 0.045 $ is in excellent agreement with the results from trajectory simulations, substantiating the validity of the Gaussian approximation.
This shows that the Gaussian action is a valid approximation to capture the whole statistics of quantum self-closing trajectories, and this might be valuable in any realization in which identifying the most probable trajectories might not be feasible.

\section{Conclusions}
In this work, we have developed an action formalism that incorporates an overall phase for the dynamics of a single qubit subject to Gaussian measurements based on the formalism developed by Chantasri–Dressel–Jordan. 
We have utilized this formalism to define a suitable Lagrangian density with associated holonomic constraints. 
The inclusion of a phase degree of freedom allows us to determine the statistical properties of measurement-induced geometric phases for both trajectories with open boundary conditions (open geometric phase) and sets of self-closing trajectories (closed geometric phase). 
We determined the topological properties of the open geometric phase for most likely trajectories and a variety of initial conditions, showing that transition in a topological number is 
a general feature, although the quantitative values of the critical measurement are protocol-dependent.
Most importantly, the formalism allows us to study self-closing geometric phases, for which we show that a transition in the topological number is present for the entire Bloch sphere, although the underlying mechanism differed from that of open-geometric phases: in the former, the transition is dictated by the competition of local most-likely trajectories, while in the latter, the geometric phase is unobservable (occurring with vanishing probabilities) at the transition point. 
Furthermore, we have shown that including multiple trajectories around the optimal one via Gaussian approximation is an excellent approximation of the full distribution simulated numerically, and leads to modifications of the transition critical point. 

The formalism developed in this work can be the basis for the efficient study of measurement-induced dynamics in more complex systems, including multiple qubits and their entanglement dynamics. It can also be exploited for a more direct connection of self-closed geometric phases to experiments, where the statistics of multiple trajectories are unavoidable.

\section*{Acknowledgments} 
We would like to thank Y. Gefen and J. Ruostekoski for helpful discussions. A.R. acknowledges support from the Royal Society,
grant no. IECR2212041

\bibliographystyle{unsrt.bst}
\bibliography{apssamp}
\end{document}